\documentclass[twocolumn,preprintnumbers,amsmath,amssymb,superscriptaddress]{revtex4}
\usepackage{graphicx}
\usepackage{dcolumn}
\usepackage{bm}
\usepackage{soul}
\usepackage{color}
\usepackage{epstopdf}
\usepackage[version=3]{mhchem}
\usepackage{lipsum}
\usepackage[outercaption]{sidecap}
\usepackage{floatrow}
\usepackage{hyperref}

\begin{document}

\title{Application of the Statistical Moment Method to Melting Properties of Ternary Alloys with FCC Structure}
\author{Tran Dinh Cuong}
\email{cuong.trandinh@phenikaa-uni.edu.vn}
\affiliation{Hanoi National University of Education, 136 Xuan Thuy, Cau Giay, Hanoi 100000, Vietnam}%
\affiliation{Faculty of Materials Science and Engineering, Phenikaa Institute for Advanced Study, Phenikaa University, Hanoi 100000, Vietnam}
\affiliation{Faculty of Information Technology, Artificial Intelligence Laboratory,  Phenikaa University, Hanoi 100000, Vietnam}%
\author{Nguyen Quang Hoc}
\affiliation{Hanoi National University of Education, 136 Xuan Thuy, Cau Giay, Hanoi 100000, Vietnam}
\author{Anh D. Phan}
\email{anh.phanduc@phenikaa-uni.edu.vn}
\affiliation{Faculty of Materials Science and Engineering, Phenikaa Institute for Advanced Study, Phenikaa University, Hanoi 100000, Vietnam}
\affiliation{Faculty of Information Technology, Artificial Intelligence Laboratory,  Phenikaa University, Hanoi 100000, Vietnam}
\date{\today}

\date{\today}

\begin{abstract}
The high-pressure melting properties of the ternary alloy AlCuSi with face-centred cubic structure is theoretically investigated using the statistical moment method. We calculate the melting temperature for the alloy under pressure up to 80 GPa. The dependence of the melting temperature on the content of alloying elements is also studied. Our results agree well with previous experiments, simulations, and other theoretical calculations. 
\end{abstract}

\maketitle


\section{Introduction}
Aluminum alloys have been intensively investigated due to their wide range of applications in transportation, automotive and aerospace industries. The alloys have high specific strength, good corrosion resistance, and are lightweight \cite{1,2,3}, while the cost is relatively low. A pure aluminum material is a light metal that avoids the progressive oxidation process found in steel. Instead, an inert, aluminum oxide layer is formed, which protects the material from further oxidation and corrosion due to the environment. However, like other metals, aluminum has a relatively low strength (approximately 90 MPa). Appropriately adding elements to aluminum does not only improve the mechanical strength, but also enhances other inherent properties. Copper and silicon are typical choices for alloying with aluminum in this perspective. Interestingly, copper strengthens the modulus of pure aluminum and Al-Si alloys but it reduces the corrosion resistance of Al. Meanwhile, the presence of silicon in Al-Si alloys reduces cracking, minimizes shrinkage porosity, improves fluidity and leaves the material lightweight. Employing lightweight aluminum alloys reduces fuel consumption in engines significantly, which decreases pollution and greenhouse gases. Consequently, combining Al, Cu, and Si has considerably improved the desirable, synergistic and complementary effects. Despite various theoretical and experimental studies, the ternary system AlCuSi has still not been fully understood.

Determining the melting temperature of metal, particularly aluminum, and its alloys under extreme pressure has been a topical question in condensed-matter physics, astrophysics, and geophysics because of numerous applications in high-pressure science and technology \cite{6,40,41,42}. Two main experimental techniques based on direct and indirect measurements have been proposed to explore the high-pressure melting behaviors: diamond-anvil cell (DAC) technique \cite{4,16} and shock-wave induced melting \cite{5,17}. Although the approaches can provide the pressure dependence of the melting up to hundreds of GPa, they have their own limitation \cite{8}. Sometimes these experimental results disagree with each other and even simulations \cite{18}. Using classical molecular dynamic (CMD) simulation or ab initio molecular dynamic simulation (AIMD) is very time consuming, and the problem becomes more complicated when materials contain impurities \cite{100}. Most computational studies performed by CMD and AIMD only consider the high-pressure melting temperature at a few concentrations of impurities \cite{19}. Another computational method, so-called the CALculation of PHAse Diagrams (CAPHAD) method \cite{20}, based on thermophysical properties and phase behaviours of alloys overcomes the shortcomings. In the CALPHAD method, all experimental and theoretical data on thermodynamic and phase equilibria in the considered system must be first collected and analyzed. Then some adjustable parameters are proposed for the Gibbs free energy so that one can recalculate all obtained information of each phase. From these, thermodynamic properties in unexplored regions can be reliably predicted. Although using the CALPHAD approach requires a set of systematic experimental data and integrates various softwares, we can gain insights into a whole physical picture of how the thermodynamic properties of a multi-component crystal depend on impurities. However, the fact that the CALPHAD method only calculates the phase diagrams at certain pressures, means that the melting information of the crystal under pressure obtained from the CALPHAD approach is discrete. Consequently, it is essential to develop theoretical models to identify decisive factors in the melting phenomenon.

From a theoretical point of view, the statistical moment (SMM) method provides mathematically simple but comprehensive description of the mechanical and thermodynamic quantities of crystalline materials. The method constructs the quantum density matrix associated with the anharmonicity and correlative effects of all moments in the systems. To determine the melting temperature, the SMM method employs the absolute stability limit of the crystalline state \cite{10}.  By following this path, it is easy to consider the effects of pressure and impurities on the melting properties of the crystals without heavy computation workloads. Theoretical results performed by the SMM approach have agreed well with experiments in many systems \cite{11,12}. However, in previous studies, authors have only applied the SMM method to predict the melting of metals and binary alloys \cite{10,11,12}.

In this paper we propose, for the first time, an extension of the statistical moment method to predict the melting temperature of ternary AlCuSi alloys having face-centred cubic (fcc) structure under pressure up to 80 GPa. The weight percent of Cu and Si in the alloy range from zero to the maximum values for which our systems still exists in alloy form. Our theoretical calculations are also compared to prior experimental and simulation results.

\begin{figure}[htp]
\includegraphics[width=8.5cm]{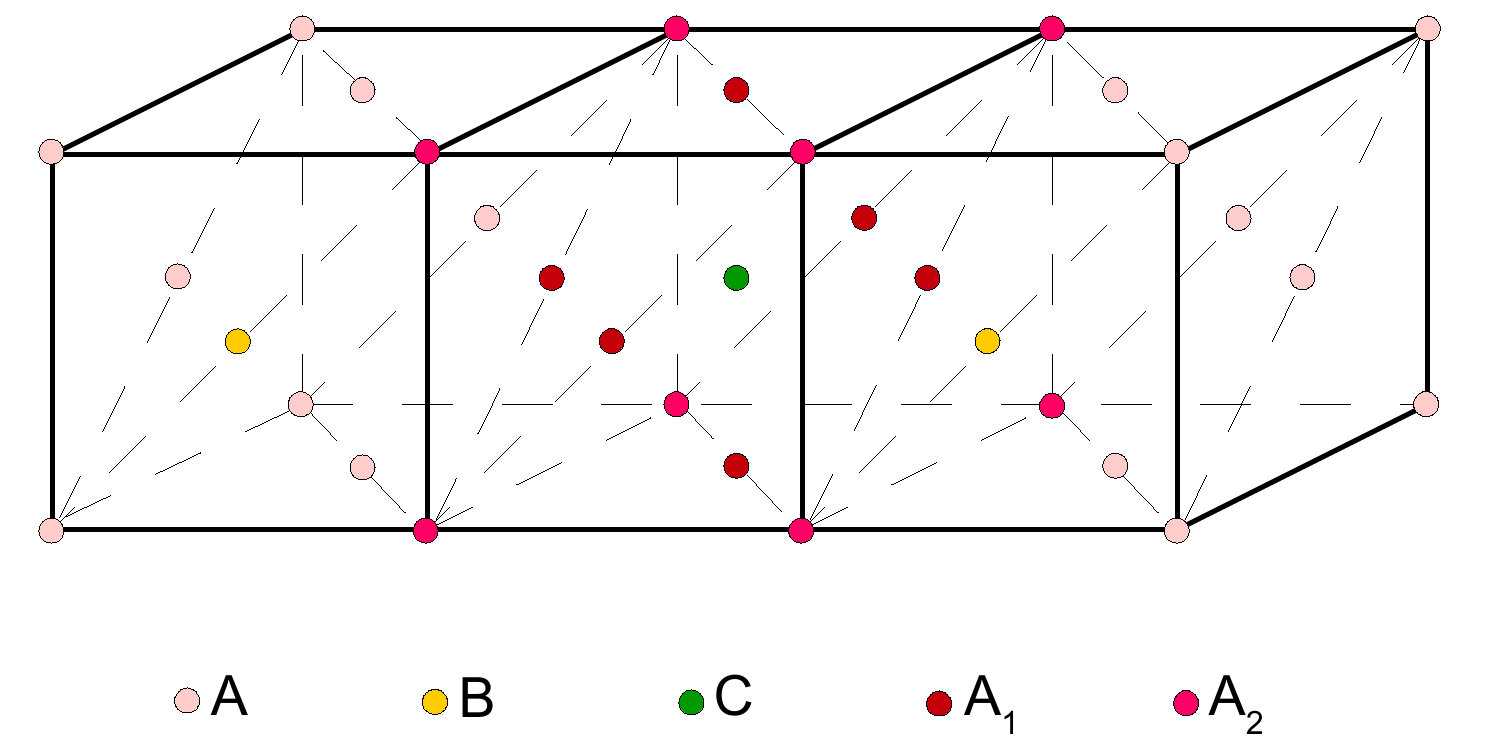}
\caption{\label{fig:1}(Color online) Illustration of crystallographic structures for ternary fcc alloys.}
\end{figure}

\section{THEORETICAL BACKGROUND}
A schematic illustration of ternary fcc alloys with the general formula ABC is shown in Fig. \ref{fig:1}, where $A$ is a main element, $B$ is a substitutional atom, and $C$ is an interstitial atom. Since the SMM method requires the symmetry of the crystal lattice, we assume that the substitutional atom $B$ is located at the centre of the cubic face and the interstitial atom $C$ is at the centre of the cube. Additionally, the short-range interaction between atoms allows us to suppose that the distance between the atom $B$ and $C$ (as depicted in Fig.\ref{fig:1}) is sufficiently long to ignore the $B-C$ interaction. We only consider the interaction of the atom $C$ with the main atoms $A$, which are closest to it ($A_1$ and $A_2$). In a previous work \cite{9}, we derived the analytic expressions for the nearest neighbor distance $a_X$, the cohesive energy $u_{0X}$ and the alloy parameters $k_X, \gamma_{X}, \gamma_{1X}, \gamma_{2X}$  for the atom $X$ ($X = A, B, C, A_1, A_2$). Here $k_X$, $\gamma_{1X}$, and $\gamma_{2X}$ are the harmonic force constant, the first and second anharmonic terms calculated using the Taylor expansion of the cohesive energy $u_{0X}$, respectively, and $\gamma_{X} = 4(\gamma_{1X}+\gamma_{2X})$. The atomic concentration $c_X$  is determined by

\begin{eqnarray}
c_B = \frac{N_B}{N}, \quad c_C=\frac{N_C}{N} &,& \quad c_{A_1}=  6c_C, \quad c_{A_2}=  8c_C, \nonumber\\ c_A &=& 1- c_B - 15c_C,
\label{eq:1}
\end{eqnarray}
where $N_X$ is the number of atoms $X$, and $N$ is the total atomic number of the system.

The mean nearest neighbor distance between two main atoms $A$ in the ternary alloy $ABC$ is approximately given by \cite{9}
\begin{eqnarray}
a_{ABC}=\frac{(1-c_B)B_{TAC}a_{AC}+c_BB_{TB}a_B}{(1-c_B)B_{TAC}+c_BB_{TB}},
\label{eq:2}
\end{eqnarray}
where $a_{AC}$ and $B_{TAC}$ are the mean nearest neighbor distance and the isothermal bulk modulus of the interstitial alloy $AC$, respectively, and $B_{TB}$ is the isothermal bulk modulus of the pure metal $B$.

The Helmholtz free energy for the ternary alloy is written by \cite{9}
\begin{eqnarray}
\Psi_{ABC} = \sum_{X}c_X\Psi_X - TS_{ABC},
\label{eq:3}
\end{eqnarray}
where $\Psi_X$ is the free energy of the atom $X$, and $S_{ABC}$ is the configurational entropy. The free energy $\Psi_X$ is approximately expressed by 

\begin{eqnarray}
\Psi_{X} &\approx& \frac{N}{2}u_{0X}+3N\theta [x_X+\ln(1-e^{-2x_X})] \nonumber\\
&+& 3N\left(\frac{\theta^2}{k_X^2}\left[\gamma_{2X}X_X^2-\frac{2\gamma_{1X}}{3}\left(1+\frac{X_X}{2}\right) \right] \right.\nonumber\\
&+& \left.\frac{\theta^3(2+X_X)}{k_X^4}\right.\nonumber\\ 
&\times& \left. \left[\frac{4\gamma_{2X}^2X_X}{3} - 2(\gamma_{1X}^2+2\gamma_{1X}\gamma_{2X})(1+X_X)\right] \right), \nonumber\\
 S_{ABC} &=& k_{0B}\ln\left(\frac{N!}{N_A!N_B!N_C!N_{A_1}!N_{A_2}!} \right), 
\label{eq:4}
\end{eqnarray}
where $T$ is the temperature, $k_{0B}$ is the Boltzmann constant, $m_X$ is the atomic mass of the atom $X$, $\theta = k_{0B}T$ is the thermal energy, $\omega_X$ is the oscillation frequency of atoms X around their equilibrium position, $x_X =\hbar\omega_X/2\theta$ is the harmonic energy normalized by the thermal energy, and $X_X = x_X\coth{x_X}$ is a dimensionless quantity to simplify our analytical expression of the free energy $\Psi_X$. From this, we can obtain the equation of state of the ternary alloy ABC as follows,
\begin{eqnarray}
P =-\frac{1}{N}\left(\frac{\partial\Psi_{ABC}}{\partial \nu_{ABC}} \right)_T =  \frac{3\gamma_G\theta}{\nu_{ABC}} -\frac{a_{ABC}}{6\nu_{ABC}}\sum_{X}c_X\frac{\partial u_{0X}}{\partial a_X},
\label{eq:5}
\end{eqnarray}
where $\nu_{ABC} =a_{ABC}^3/\sqrt{2}$ is the volume of a fcc unit cell and $\gamma_G$ is the Gruneisen parameter
\begin{eqnarray}
\gamma_G = -\frac{a_{ABC}}{6}\sum_X\frac{c_X}{k_X}\frac{\partial k_X}{\partial a_X}X_X.
\label{eq:6}
\end{eqnarray}

Now, solving Eq. (\ref{eq:5}) and (\ref{eq:6}) under the condition of the absolute stability limit of the crystalline state, $\left(\partial P/\partial\nu_{ABC} \right)_{T=T_S} = 0$, gives us  the critical temperature $T_S$ for the stable crystal structure,
\begin{widetext}
\begin{eqnarray}
T_S = \cfrac{2P\nu_{ABC} + \cfrac{a_{ABC}^2}{6}\sum_Xc_X\cfrac{\partial^2 u_{0X}}{\partial a_X^2}-\cfrac{\hbar a_{ABC}^2}{4}\sum_X\cfrac{c_X\omega_X}{k_X}\left[\cfrac{1}{2k_X}\left(\cfrac{\partial k_X}{\partial a_X} \right)^2 - \cfrac{\partial^2 k_X}{\partial a_X^2} \right] }{\cfrac{k_{0B}a_{ABC}^2}{4}\sum_X\cfrac{c_X}{k_X^2}\left(\cfrac{\partial k_X}{\partial a_X} \right)^2}.
\label{eq:7}
\end{eqnarray}
\end{widetext}

The critical temperature $T_S$ has a strong correlation with the melting temperature $T_m$. For metals, authors in Ref.\cite{10,11} assumed that $T_m \approx T_S$ and the assumption provides a good agreement between SMM calculations and experimental data. When applying the SMM approach to the melting of binary alloys while assuming that $T_m \approx T_S$, the theoretical predictions exhibit a significant deviation from experiments \cite{35}. Thus, authors in Ref. \cite{12} calibrated the SMM results by using the melting temperature at zero pressure from the experimental data to obtain consistency between the theory and experiment. Consequently, the presence of two additional compositions in ternary alloys is expected to increase the difference between theory and experiment. 

In the present study, we introduce a correction to the estimated $T_m$ from $T_S$. Equation (\ref{eq:5}) shows that the temperature $T$ is known as a function of the nearest neighbor distance $a_{ABC}$ at a given pressure: $T = f(a_{ABC})$. When the difference between $T_S$ and $T_m$ is supposed to be small, $a_{ABC}(P,T_S)$ is also close to $a_{ABC}(P,T_m)$. This assumption induces a possibility of using a Taylor expansion for $T_m$  around $a_{ABC}(P,T_S)$. Then, we have
\begin{widetext}
\begin{eqnarray}
T_m \approx T_S + \frac{a_{ABC}(P,T_m)-a_{ABC}(P,T_S)}{k_{0B}\gamma_G}\left(\frac{P\nu_{ABC}}{a_{ABC}(P,T_S)} + \sum_X\frac{c_X}{18}\left[\left(\frac{\partial u_{0X}}{\partial a_{X}} \right)_{T=T_S} + a_{ABC}(P,T_S)\left(\frac{\partial^2 u_{0X}}{\partial a_{X}^2} \right)_{T=T_S} \right] \right).
\label{eq:8}
\end{eqnarray}
\end{widetext}

In principle, we can obtain the melting curve of a crystal by solving numerically Eqs. (\ref{eq:7}) and (\ref{eq:8}). But if we know the melting temperature $T_m(0)$ at zero pressure, it is easier to determine the melting temperature $T_m(P)$ at pressure $P$ \cite{13},
\begin{eqnarray}
T_m(P) = \frac{T_m(0)B_0^{1/B_0'}}{G(0)}\frac{G(P)}{\left(B_0 + B_0'P \right)^{1/B_0'}},
\label{eq:9}
\end{eqnarray}
where $G(P)$ and $G(0)$ are the shear moduli at pressure $P$ and zero pressure, respectively, $B_0$ is the isothermal elastic modulus at zero pressure, $B_0' = \left(dB_T/dP \right)_{P=0}$, and $B_T$ is the isothermal elastic modulus at pressure $P$. It is important to note that composites are theoretically supposed to undergo a homogeneous melting mechanism in our theoretical approach.

\section{NUMERICAL RESULTS AND DISCUSSION}
Finding analytic interatomic potential functions has been a challenging problem in quantum mechanics. But by combining theory and experiment, researchers have found a variety of empirical potentials to determine effectively the macroscopic properties of solids, such as the Mie-Lennard-Jones potential, the Morse potential, and other potentials of the embedded atom method and the modified embedded atom method. The empirical potential parameters are calculated by fitting with experimental data, so the predictions from the empirical potential could more reliable than using DFT potential. In our previous work, we indicated that the Mie-Lennard-Jones potential is the most efficient empirical potential to capture the physical nature of the problem, while keeping flexible adjustable parameters. Furthermore, in many cases the results obtained by using the Mie-Lennard-Jones potential for the SMM’s approach provide a better agreement with experimental measurements than other potentials.  Consequently, we continue to use the Mie-Lennard-Jones potential written in Eq.(\ref{eq:10}) to describe the interaction between atoms in the ternary alloys AlCuSi.

\begin{eqnarray}
\varphi(r) = \frac{D}{n-m}\left[m\left(\frac{r_0}{r} \right)^n-n\left(\frac{r_0}{r} \right)^m \right],
\label{eq:10}
\end{eqnarray}
where $D$, $n$, $m$, and $r_0$ are Mie-Lennard-Jones parameters for Al-Al and Cu-Cu potentials taken from Ref.\cite{14}, while the parameters for the Si-Si interaction are given in Ref. \cite{15}. Since the data for the interaction between Al and Cu atoms is unavailable, the Al-Cu potential can be achieved by averaging the Al-Al and Cu-Cu potentials 
\begin{eqnarray}
\varphi_{Al-Cu}(r) = \frac{\varphi_{Al-Al}(r) +\varphi_{Cu-Cu}(r)}{2}.
\label{eq:11}
\end{eqnarray}

To determine the Al-Si potential, we use the Berthelot-Good-Hope combining rule as follows \cite{36} 
\begin{eqnarray}
D = \sqrt{D_{Si-Si}D_{Al-Al}}, \quad r_0 = \sqrt{r_{0,Si-Si}r_{0,Al-Al}},
\label{eq:12}
\end{eqnarray}
then parameters $m$ and $n$ are adjusted to be consistent with experiments. The interatomic-potential parameters used in this work are listed in Table \ref{table:1}

\begin{table}[htp]
\caption{\label{tab:table1} Parameters for the Mie-Lennard-Jones potentials between Al-Al, Cu-Cu, Si-Si, and Al-Si.}
\begin{center}
\begin{tabular}{|c|c|c|c|c|}
\hline
 & $D$ ($eV$) & $n$ & $m$ &  $r_0$ ($\AA$)  \\
\hline
Al-Al & 0.2580  & 11.0 & 5.5 & 2.85 \\
\hline
Cu-Cu & 0.2929 & 11.0 & 5.5 & 2.55  \\
\hline
Si-Si & 2.32 & 4.0 & 2.48 & 2.35\\
\hline
Al-Si & 0.7737 & 8.5 & 4.0 & 2.59\\
\hline
\end{tabular}
\end{center}
\label{table:1}
\end {table}

\begin{figure}[htp]
\includegraphics[width=8.5cm]{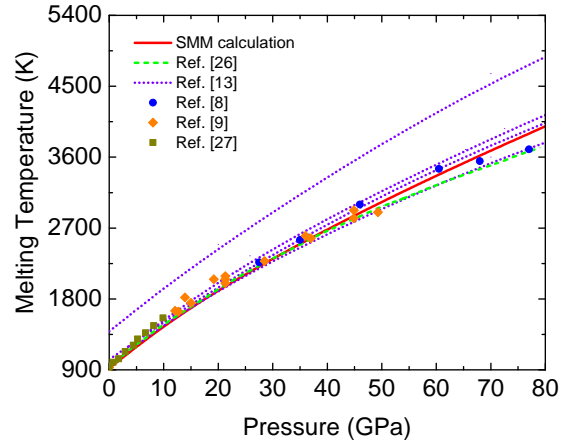}
\caption{\label{fig:2}(Color online) The melting temperature of Al as a function of applied pressure calculated by the SMM method, simulations \cite{18,7}, and experiments \cite{4,16,21}.}
\end{figure}

First of all, to validate the SMM approach and our chosen parameters, we investigate the melting curve of Al without impurities. Figure \ref{fig:2} shows how the melting temperature of Al given by our theoretical calculations, previous simulations \cite{18,7}, and experiments \cite{4,16,21} depends on pressure. The SMM approach exhibits a quantitatively good prediction with prior works. When the pressure increases from 0 to 80 GPa, the melting temperature $T_m$ of Al grows monotonically from 921 to 3992 $K$. The shrinkage of the lattice distortion under compression suppresses the volumetric expansion and increases the cohesive energy. Consequently, the substance requires more thermal energy to melt.  

\begin{figure}[htp]
\includegraphics[width=8.5cm]{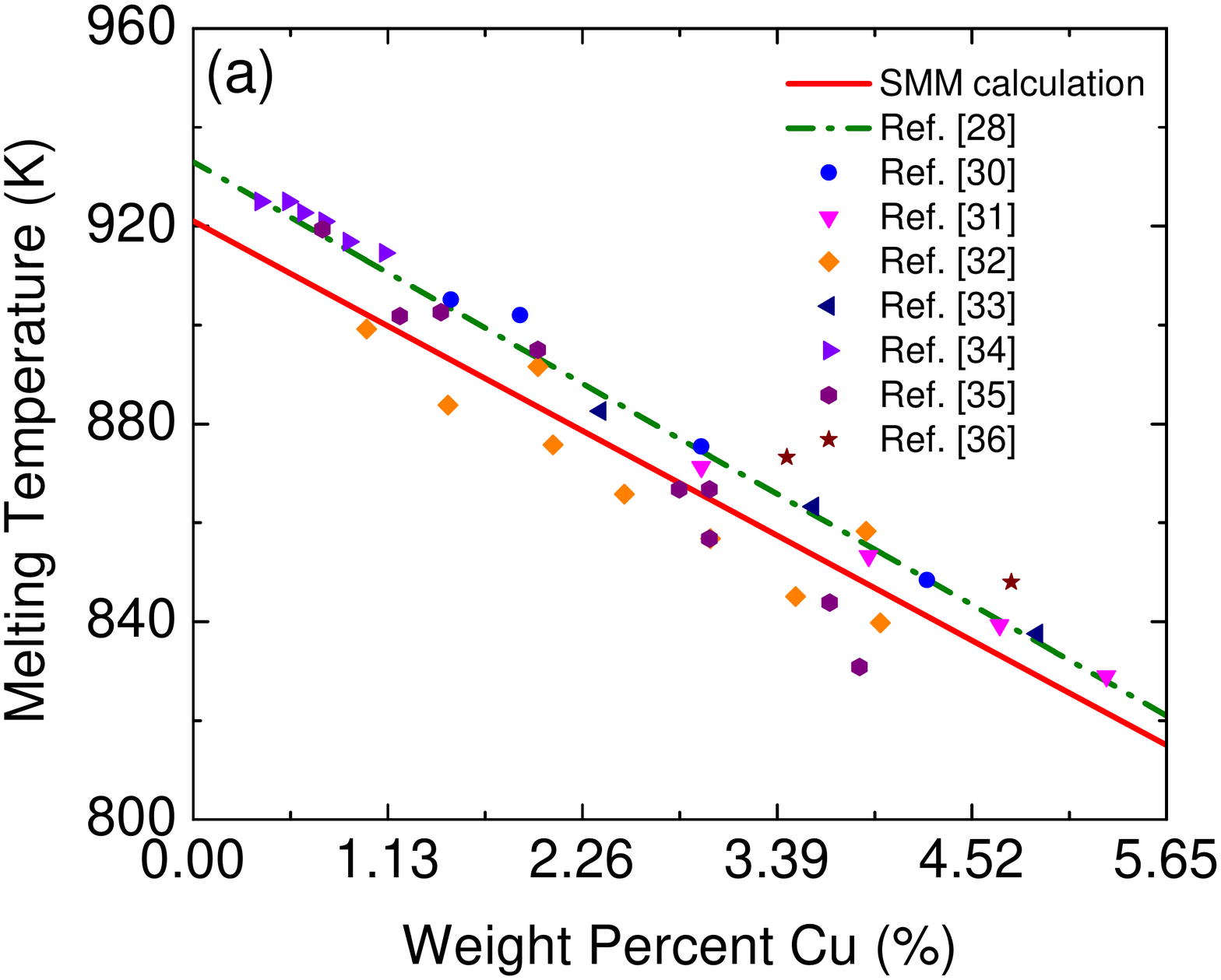}
\includegraphics[width=8.5cm]{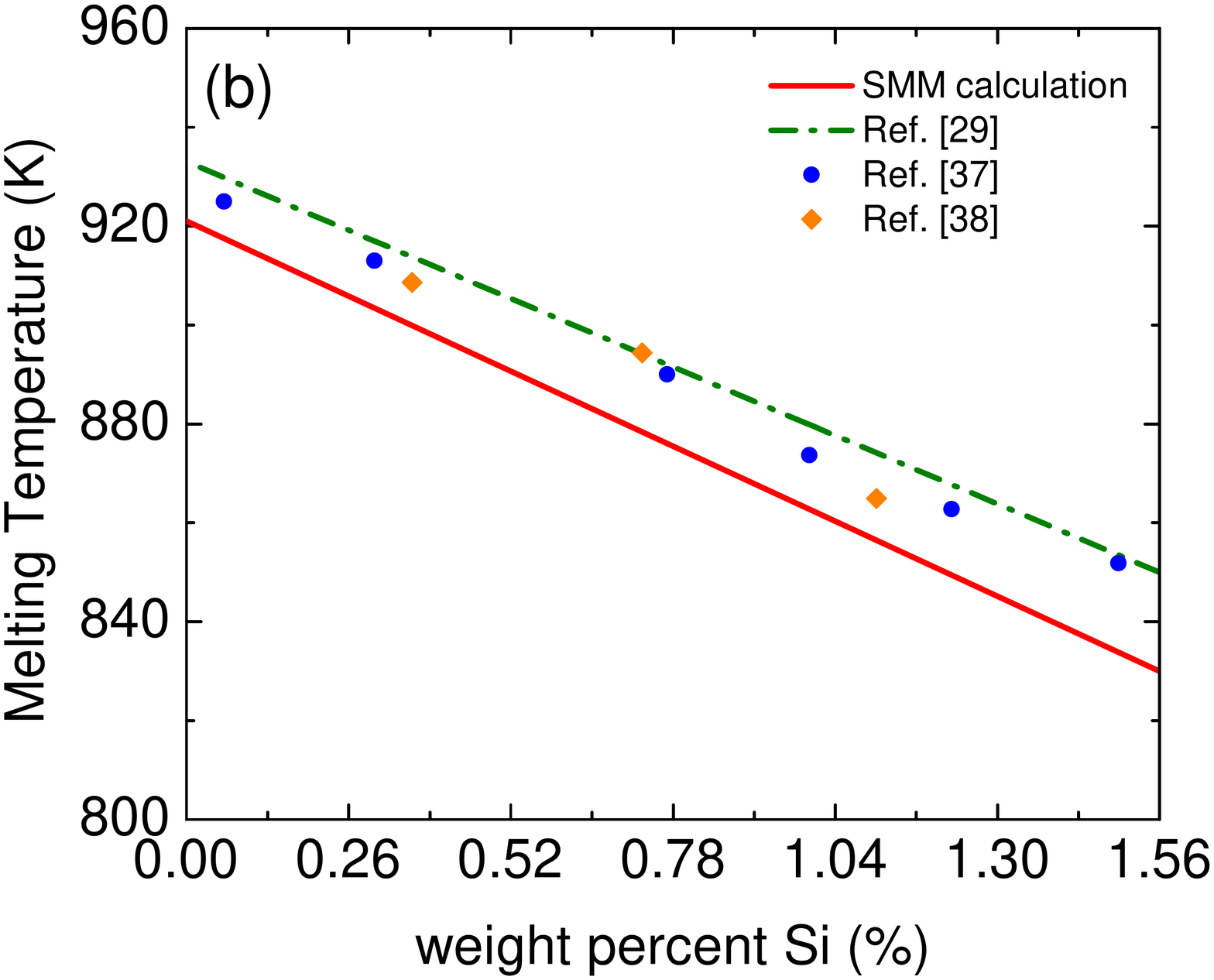}
\caption{\label{fig:3}(Color online) The melting temperature of binary aluminum alloy at zero pressure as a function of (a) copper content and (b) silicon content, calculated by the SMM method, CALPHAD simulation \cite{22,23}, and experiments \cite{24,25,26,27,28,29,30,31,32}.}
\end{figure}

Next, we consider influences of alloying elements on the melting properties of binary aluminum alloys. Effects of the substitutional element Cu and the interstitial element Si on the melting temperature of Al at zero pressure are shown in Figs. \ref{fig:3}a and \ref{fig:3}b, respectively. The melting temperature decreases with an increase of alloying element content. Our theoretical calculations are in excellent accordance with previous studies \cite{22,23,24,25,26,27,28,29,30,31,32}. The average melting slopes ($\Delta T/\Delta w_{Cu}$) of the aluminum-copper alloy calculated using our analysis and CALPHAD simulation \cite{22} are about -18.8 $K/\%$ and -19.8 $K/\%$, respectively. Here $w_X$ denotes the weight percentage of element $X$. Similarly, while CALPHAD simulation \cite{23} predicts the melting temperature to decrease with silicon content with a melting slope of -53.2 $K/\%$, a slope of the melting curve predicted using the SMM method is -58.3 $K/\%$. These findings suggest that adding silicon to aluminum reduces the melting temperature more significantly than adding copper at a given concentration of the alloying element. Physically, since copper and aluminum have different lattice constants, the addition of copper alloying element in aluminum causes inhomogeneous lattice distortion and enhances solid-state disorder. Atoms are already dislocated from their positions in the pure form. Thus, the atoms require less thermal energy to melt and the melting temperature of aluminum alloys is lowered in comparison with the pure counterpart. However, Al and Cu have many structural similarities so that substituting Cu into Al does not deform much the crystal lattice. Meanwhile, the presence of the interstitial element Si substantially distorts the lattice structure of the main component Al. Consequently, the mechanical and thermodynamic properties of the crystal are remarkably changed, particularly for the mean nearest neighbor distance and the melting temperature.

\begin{figure}[htp]
\includegraphics[width=8.5cm]{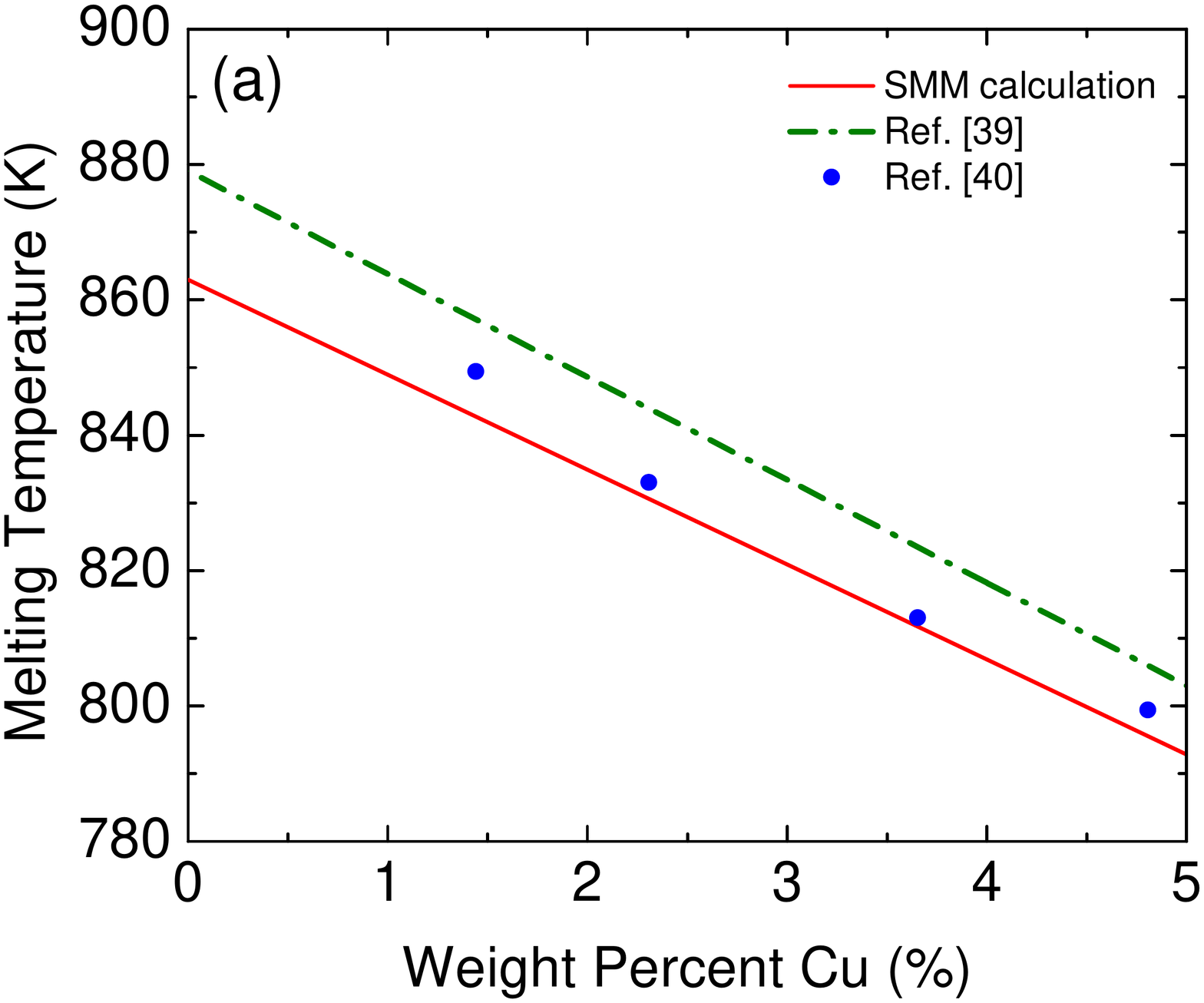}
\includegraphics[width=8.5cm]{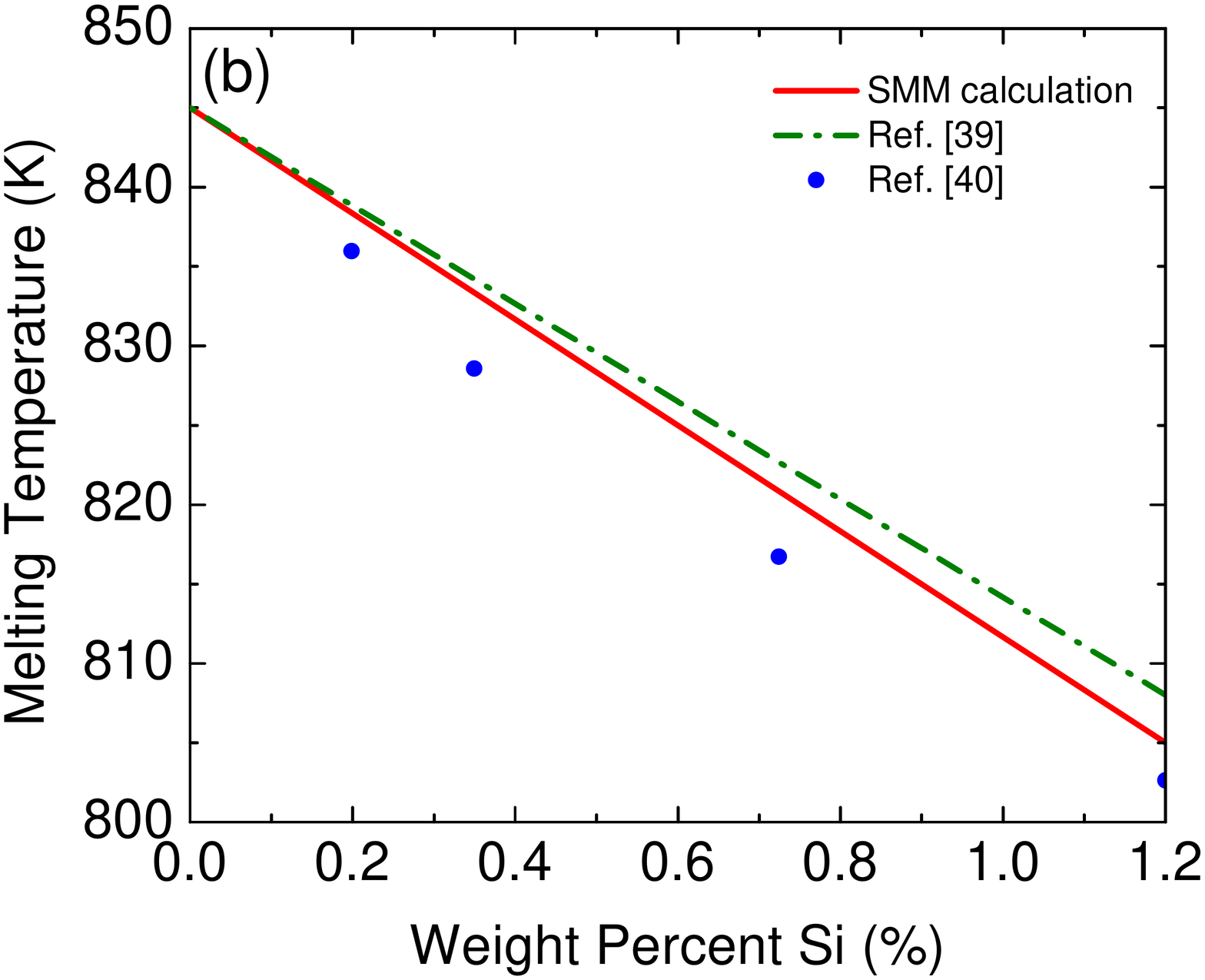}
\caption{\label{fig:4}(Color online) The melting curves of (a) alloy \ce{AlCuSi_{1\%}}, and (b) alloy \ce{AlCu_{4\%}Si} (b) at zero pressure obtained from the SMM method, CALPHAD simulation \cite{33}, and experiments \cite{34} as a function of Cu and Si content.}
\end{figure}

\begin{figure}[htp]
\includegraphics[width=8.5cm]{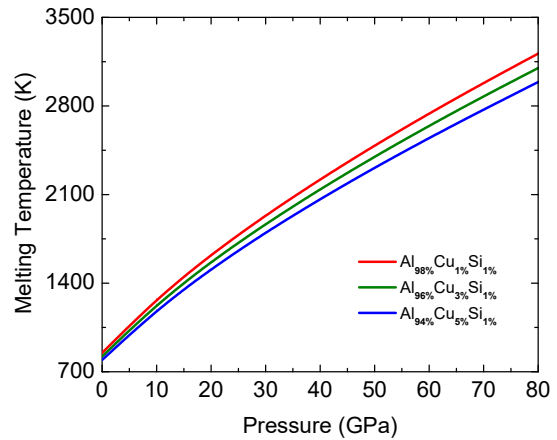}
\caption{\label{fig:5}(Color online) The melting curve of ternary AlCuSi alloys under pressure up to 80 GPa calculated by the SMM method.}
\end{figure}

Figure \ref{fig:4} shows how the melting temperature of ternary alloy \ce{AlCuSi_{1\%}} (Fig. \ref{fig:4}a) and \ce{AlCu_{4\%}Si} (Fig. \ref{fig:4}b) at zero pressure depend on Cu and Si content, respectively, calculated using the SMM method, CALPHAD simulation \cite{33}, and experiments \cite{34}. The SMM approach and CALPHAD simulation \cite{33} give the average melting slope of alloy \ce{AlCuSi_{1\%}} of approximately -14 $K/\%$ and -15.2 $K/\%$, respectively. For the ternary alloy \ce{AlCu_{4\%}Si}, the average slope of the melting curve predicted by CALPHAD simulation is -30.8 $K/\%$, which is comparable to our SMM analysis (-33.3 $K/\%$). Although both our SMM-based calculations and the CALPHAD simulation \cite{33} agree well with prior experiments \cite{34}, our results are relatively closer to experimental data than the simulation. Our theoretical studies in Figs. \ref{fig:3} and \ref{fig:4} imply that the chosen interatomic potentials in Eq.(\ref{eq:10}), (\ref{eq:11}), and (\ref{eq:12}) are reliable and can be used to predict quantitatively the high-pressure melting curve of the ternary alloy AlCuSi.

Similar to the melting curve of pure aluminum under compression in Fig. \ref{fig:2}, the melting temperature of AlCuSi alloys approximately increases by a factor of 4 with pressures up to 80 GPa as shown in Fig.\ref{fig:5}. The melting curves seem to be unaffected by a variation of Cu content from 1 to 5$\%$. Figure \ref{fig:4} reveals that the presence of either Cu or Si in an aluminum material enhances the melting process. Thus, the melting curves of ternary alloys are expected to be lower than that of pure aluminum.

A Simon analytic expression relating the melting temperature and pressure, which is often used to fit data, is
\begin{eqnarray}
T_m = T_{m0}\left(\frac{P}{P_0}+1\right)^b,
\end{eqnarray}
where $T_{m0}$, $P_0$, and $b$ are adjustable fit parameters. The calculated parameters for Al and its studied ternary systems are listed in Table \ref{table:2}

\begin{table*}[htp]
\caption{\label{tab:table1} The parameters obtained fitting our numerical results for Al and its ternary systems under compression ranging from 0 to 80 GPa to a Simon equation.}
\begin{center}
\begin{tabular}{|c|c|c|c|c|}
\hline
 & Al & $Al_{98\%}Cu_{1\%}Si_{1\%}$ & $Al_{96\%}Cu_{3\%}Si_{1\%}$ & $Al_{94\%}Cu_{5\%}Si_{1\%}$  \\
\hline
$T_{m0}$ (K) & 924.66  & 851.98 & 823.83 & 795.88 \\
\hline
$P_0$ (GPa) & 10.4 & 11.44 & 11.51 & 11.63  \\
\hline
$b$ & 0.6759 & 0.6383 & 0.6386 & 0.6404\\
\hline
\end{tabular}
\end{center}
\label{table:2}
\end {table*}

\section{CONCLUSION}
We have presented a simple but very effective approach, the statistical moment method, to calculate the high-pressure melting curve of ternary alloys with fcc structure. The theoretical results have been performed for pure aluminum and aluminum alloys using Mie-Lennard-Jones pair potentials in the interval of pressure from 0 to 80 GPa. The dependence of the melting temperature of pure aluminum on compression is in a strong agreement with previous simulations and experiments. At zero pressure, an addition of Cu or Si to aluminum alloys reduces the melting temperature. For the same content of alloying elements, the silicon component shows a greater melting point reduction than copper. Our numerical results are in a good agreement with experiments, ab initio, and CALPHAD simulations. Finally, based on success of the proposed approach, we predict the melting curve of ternary AlCuSi alloys.

\begin{acknowledgments}
This research is funded by Vietnam National Foundation for Science and Technology Development (NAFOSTED) under grant number 103.01-2017.63.
\end{acknowledgments}

\end{document}